\documentclass[conference]{IEEEtran}
\IEEEoverridecommandlockouts
\usepackage{cite}
\usepackage{diagbox}
\usepackage{amsmath,amssymb,amsfonts}
\usepackage{algorithmic}
\usepackage{verbatim}
\usepackage{algorithmic}
\usepackage{graphicx}
\usepackage{textcomp}
\usepackage{xcolor}
\usepackage{multirow}
\usepackage{balance}
\usepackage{array}
\usepackage{tabu}
\usepackage{longtable}[=v4.13]
\usepackage{booktabs} 
\usepackage{color}
\usepackage{subfigure}
\usepackage{caption}
\usepackage{balance}
\usepackage{enumitem}
\usepackage{orcidlink}
\usepackage{tabularx}
\usepackage{physics}
\usepackage{float}
\usepackage{multicol}

\def\BibTeX{{\rm B\kern-.05em{\sc i\kern-.025em b}\kern-.08em
    T\kern-.1667em\lower.7ex\hbox{E}\kern-.125emX}}

\begin{document}

\title{Hardware-Efficient CNNs: Interleaved Approximate FP32 Multipliers for Kernel Computation}
\author{
\IEEEauthorblockN{Bindu G Gowda \orcidlink{0000-0003-2797-2363}}
\IEEEauthorblockA{IIIT-Bangalore, India\\
\textit{bindu.gowda@iiitb.ac.in}
}
\and
\IEEEauthorblockN{Yogesh Goyal}
\IEEEauthorblockA{IIIT-Bangalore, India\\
\textit{yogesh.goyal@iiitb.ac.in}
}
\and
\IEEEauthorblockN{Yash Gupta}
\IEEEauthorblockA{IIIT-Bangalore, India\\
\textit{yash.gupta514@iiitb.ac.in}
}
\and
\IEEEauthorblockN{Madhav Rao \orcidlink{0000-0003-2278-9148}}
\IEEEauthorblockA{IIIT-Bangalore, India\\
\textit{mr@iiitb.ac.in}
}
}

\maketitle

\begin{abstract}
Single-precision floating point (FP32) data format, defined by the IEEE 754 standard, is widely employed in scientific computing, signal processing, and deep learning training, where precision is critical. However, FP32 multiplication is computationally expensive and requires complex hardware, especially for precisely handling mantissa multiplication. 
In practical applications like neural network inference, perfect accuracy isn't always necessary—minor multiplication errors often have little impact on final accuracy. This enables trading precision for gains in area, power, and speed.
This work focuses on CNN inference using approximate FP32 multipliers, where the mantissa multiplication is approximated by employing error-variant approximate compressors, that significantly reduce hardware cost.
Furthermore, this work optimizes CNN performance by employing differently approximated FP32 multipliers and studying their impact when interleaved within the kernels across the convolutional layers. The placement and ordering of these approximate multipliers within each kernel are carefully optimized using the Non-dominated Sorting Genetic Algorithm-II, balancing the trade-off between accuracy and hardware efficiency.
\end{abstract}



\section{Introduction}

Multipliers are heavily used in computer vision applications that involve processing and interpreting images and videos. Tasks like image filtering, feature extraction, and convolution operations in deep learning-based vision models rely extensively on multiplications. Convolutional Neural Networks (CNNs), the backbone of modern vision systems, perform millions of MAC operations per inference, making multiplier efficiency vital. However, their complex logic, multipliers leads to hardware cost in terms of area, power, and critical path delay~\cite{multiplier-saket2}.
As the multipliers are considered energy-hogging modules,
the approximations in multiplier designs have attracted much attention in the past few years.
Approximation in computing elements saves hardware resources, footprint cost, and significantly improving delay.
These approximate schemes primarily target error-resilient applications like image and, to some extent, signal processing, where post-processed outcomes remain consistent despite deviations in individual computation results~\cite{appln-image2}.
The approximate designs are expected to leverage hardware resources and improve the throughput without significantly affecting the output quality. 

LUT-based approximate multipliers (AMs) are proposed in \cite{multiplier-fpga2} that are more suitable for FPGA applications and utilize ternary addition, but it was also observed that compressor-based multiplier designs were beneficial in accelerating the multiplication process; hence, later works focused on approximate compressor designs for constructing AMs to achieve hardware improvements 
\cite{compressor8}.
The concept of approximate computing has recently been extended to hardware-aware CNNs and other computer vision applications \cite{bindu-journal}.
Although the work in \cite{alwann} provided a valuable platform for evaluating CNN accuracy with newly designed multipliers, it did not discuss hardware metrics across the CNN implementation. Most recent works have primarily focused on applying the same type of AM across the convolutional layers of the network \cite{ Bindu_raghava-Design-Space-isvlsi}. State-of-the-art (SOTA) works \cite{bindu-raghava-ApproxCNN_layerwise-iccd} have explored the possibility of employing different AMs along different convolutional layers. However, these works were restricted to using the same multiplier per layer. Moreover, all these studies aimed at designing and evaluating the integer multipliers and approximate floating point multipliers have not yet been extensively explored.

While lower-bit formats are being explored for AI/ML applications, FP32 remains essential for applications requiring higher precision and stability; 
therefore, this work aims to design and characterize approximate FP32 multipliers to balance precision with hardware efficiency, using error-diluted compressors in mantissa multiplication to minimize cumulative error in iterative multiplication operations. 
It further explores the novel concept of interleaving different multiplier configurations within kernels across convolutional layers. To the best of the authors’ knowledge, this is the first study to apply such fine-grained multiplier interleaving in CNNs. The placement sequence of these multipliers is optimized using the Non-dominated Sorting Genetic Algorithm-II (NSGA-II).

\section{Approximate Single Precision Floating Point Multiplier Design}
\label{sec:apprxFP32}
The IEEE 754 standard defines the single-precision floating-point format as a 32-bit representation for processing real numbers.
It consists of a 1-bit Sign (S), an 8-bit Exponent (E), and a 23-bit Mantissa (M), in its binary representation. 
The exponent is stored using a bias of 127, allowing for the representation of both positive and negative exponents. 
The decimal equivalent value of the FP32 \textit{Normal} and \textit{Subnormal Numbers} is expressed in the Equation~\ref{eq:FP32_format}.
\begin{equation}
\label{eq:FP32_format}
FP_{32} = 
\begin{cases}
    (-1)^S \times 2^{(E-127)} \times (1.M), & \text{if}\ E\neq0 \\ \vspace{-3mm}
    \\
    (-1)^S \times 2^{(-126)} \times (0.M), & \text{if}\ E=0 \text{ \&}\ M\neq0
\end{cases}
\end{equation}

The multiplication of two floating-point numbers involves an XOR operation on the sign bits, the addition of the exponents while accounting for bias correction, and the multiplication of the mantissa bits, including the implicit leading 1 for normal numbers or an explicit 0 for subnormal numbers.

\subsection{Compressor-Based Approximate Radix-8 Modified Booth Multiplier for Mantissa Multiplication}
\label{sec:comp-based-R8_booth}

Recent studies show that approximating the partial product (PP) reduction stage using inexact compressors is the most effective among the three multiplier stages—PP generation, PP reduction, and final addition.
In \cite{bindu_PMNM}, various approximate compressors (ACs) are proposed based on metrics like error probability, mean error, error distribution, and error direction, and are categorized into positive compressors (PCs) and negative compressors (NCs). Building on this, the work introduces positive multipliers (PMs) and negative multipliers (NMs), classified into four types based on the placement of PCs and NCs-
1) Non-Interleaving (NI): A single type of AC is used across all PP reduction stages. 
2) Stage-wise Interleaving (SI): PCs and NCs are alternated stage by stage across PP reduction stages.
3) Column-wise Interleaving (CI): PCs and NCs are interleaved column-wise within the PP reduction stages.
4) Column-cum-Stage-wise Interleaving (CSI): A hybrid scheme combining both SI and CI patterns.

\begin{figure*}[tb]
    \centering
    \includegraphics[width=0.75\textwidth]{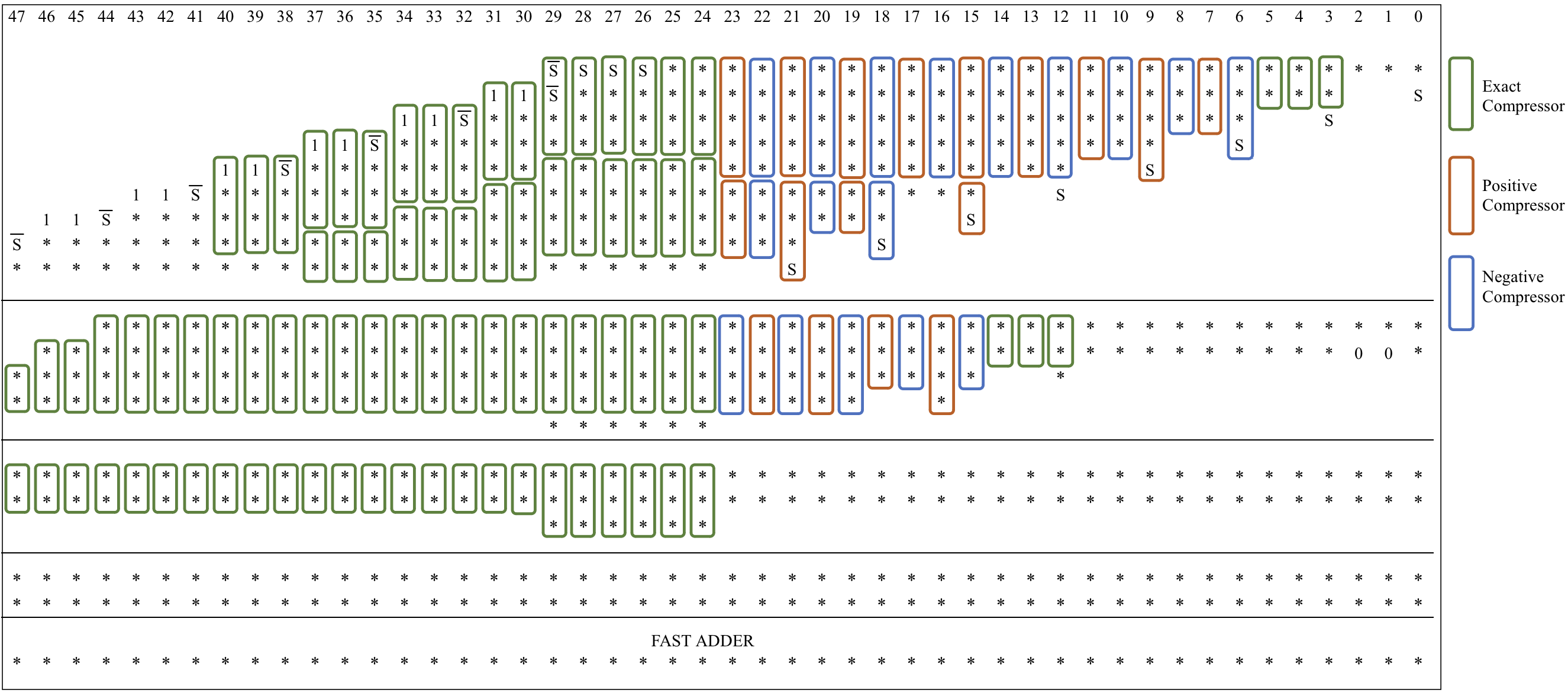}
    \caption{Dot diagram representation of $24\times24$ Radix-8 modified approximate booth multiplier with PMCSI configuration}
    \label{fig:dot_dia}
\end{figure*}

This work focuses on designing the $24 \times 24$ Radix-8 modified approximate booth multiplier (R8MABM) to handle the implicit leading 1 along with the 23-bit mantissa, with approximate compressors employed upto 24 columns of the PPs along all the reduction stages and exact compressors used in the remaining columns of PPs. 
Unlike traditional Booth multipliers that extend the sign bit of the PP row to full width of the partial product matrix (PPM), this work extends the modified PPM concept.
Figure \ref{fig:dot_dia} presents the dot diagram representation for PMCSI configuration.

\subsection{Hardware Characterization and Error Analysis}
\label{sec:FP32muls_char}
All 8 R8MABM approximate multiplier configurations were designed for mantissa multiplication, leading to the construction of 8 corresponding approximate FP32 multiplier variants.
These multipliers were implemented in Verilog HDL and synthesized in the Cadence Genus tool using the generic 45nm PDK (gpdk45) library files. The hardware characteristics are listed in Table~\ref{tab:HW_char}. In comparison with the exact FP32 multiplier, AMs with $FP32_{PMNI}$, $FP32_{PMSI}$, $FP32_{PMCI}$, $FP32_{PMCSI}$, $FP32_{NMNI}$, $FP32_{NMSI}$, $FP32_{NMCI}$, and $FP32_{NMCSI}$ configurations, exhibited the power-delay-product (PDP) benefit of 18.77\%,
23.96\%,
23.82\%,
23.94\%,
17.52\%,
24.02\%,
23.78\%, and
22.62\%, respectively. 

\begin{table}[tb]
\centering
\caption{Hardware characteristics of the Exact and Approximate FP32 Multipliers}
\label{tab:HW_char}
\resizebox{0.35\textwidth}{!}{%
\begin{tabular}{|c|c|c|c|c|} 
\hline
\begin{tabular}[c]{@{}c@{}}\textbf{Multiplier}\\\textbf{Type}\end{tabular} & \begin{tabular}[c]{@{}c@{}}\textbf{Area}\\\boldmath{$(\mu m^2)$}\end{tabular} & \begin{tabular}[c]{@{}c@{}}\textbf{Power}\\\boldmath{$(\mu W)$}\end{tabular} & \begin{tabular}[c]{@{}c@{}}\textbf{Delay}\\\boldmath{$(ps)$}\end{tabular} & \begin{tabular}[c]{@{}c@{}}\textbf{PDP}\\\boldmath{$(pJ)$}\end{tabular}  \\ 
\hline
\textbf{Exact}      & 3864.60     & 139.332        & 11966     & 1.667         \\ 
\hline
\boldmath{$FP32_{PMNI}$}       & 3627.59     & 113.623        & 11939     & 1.357         \\ 
\hline
\boldmath{$FP32_{PMSI}$}       & 3585.19     & 110.189        & 11524     & 1.270         \\ 
\hline
\boldmath{$FP32_{PMCI}$}       & 3589.29     & 108.934        & 11678     & 1.272         \\ 
\hline
\boldmath{$FP32_{PMCSI}$}      & 3594.08     & 108.736        & 11681     & 1.270         \\ 
\hline
\boldmath{$FP32_{NMNI}$}       & 3606.73     & 115.427        & 11933     & 1.377         \\ 
\hline
\boldmath{$FP32_{NMSI}$}       & 3593.05     & 109.351        & 11604     & 1.269         \\ 
\hline
\boldmath{$FP32_{NMCI}$}       & 3592.37     & 109.838        & 11588     & 1.273         \\ 
\hline
\boldmath{$FP32_{NMCSI}$}      & 3603.65     & 110.472        & 11698     & 1.292         \\
\hline
\end{tabular}}
\end{table}

Further, to evaluate the performance of the designed approximate FP32 multiplier variants, various error metrics were considered for the quantitative analysis of accuracy, reliability, and effectiveness. 
The bit-level Error Metrics including Error Rate (ER), Hamming distance (H$_d$), and Mean Absolute Bit Error (MABE); while the relative Error Metrics including Mean Relative Error (MRE), Root Mean Square Relative Error (RMSRE), and the Prediction Accuracy Metric (PRED), were evaluated in this work.
Error analysis was conducted using $N$=400000 randomly generated input combinations. 
The Hamming distance (H$_d$), which counts the differing bits between approximate and exact results, is used to compute the Mean Absolute Bit Error (MABE), indicating the average number of erroneous bits across all input combinations.
In addition, $PRED_{\tau}$ is calculated to measure the percentage of approximate results falling within an error tolerance $\tau$ relative to the exact results. This work considers a threshold of $\tau = 1$.

Table \ref{tab:err_char} presents the complete error analysis of all the AMs. 
The error rate captures only the frequency of incorrect results, not their magnitude. Even a low error rate can result in significant distortions if a few large errors occur during convolutions, potentially degrading feature maps and final model accuracy. Therefore, relying solely on error rate would be misleading, and additional metrics are considered to assess the AMs.
Among all the multipliers, $FP32_{NMNI}$ exhibited the highest MABE of 1.675 bits, implying that, on average, approximately 30.33 out of 32 bits were error-free.
Notably, despite error rates exceeding 64\%, all multipliers maintained MABE below 1.7 bits, with MRE and RMSRE in the order of $10^{-6}$ and $10^{-7}$, respectively. Additionally, all AMs achieved 99.2\% PRED, implying 99.2\% of outputs had relative errors within 1\%, indicating high accuracy. These AM designs with strategically placed PCs and NCs, effectively limit error accumulation and exhibit strong error resilience.

\begin{table}[tb]
\centering
\caption{Error Characteristics of the Exact and FP32 AMs}
\label{tab:err_char}
\resizebox{0.48\textwidth}{!}{%
\begin{tabular}{|c|c|c|c|c|c|} 
\hline
\textbf{Multiplier} & \begin{tabular}[c]{@{}c@{}}\textbf{Error rate}\\\textbf{(\%)}\end{tabular} & \textbf{MABE} & \begin{tabular}[c]{@{}c@{}}\textbf{MRE}\\\boldmath{$(10^{-6})$}\end{tabular} & \begin{tabular}[c]{@{}c@{}}\textbf{RMSRE}\\\boldmath{$(\times10^{-7})$}\end{tabular} & \begin{tabular}[c]{@{}c@{}}\boldmath{$PRED_1$}\\\textbf{(\%)}\end{tabular}  \\ 
\hline
\boldmath{$FP32_{PMNI}$}       & 72.147     & 1.574         & 6.311     & 1.156      & 99.20     \\ 
\hline
\boldmath{$FP32_{PMSI}$}       & 72.966     & 1.516         & -3.365     & 1.238      & 99.20     \\ 
\hline
\boldmath{$FP32_{PMCI}$}       & 67.471     & 1.383         & 1.364     & 1.017      & 99.20     \\ 
\hline
\boldmath{$FP32_{PMCSI}$}      & 69.904     & 1.522         & -2.425     & 1.146      & 99.20     \\ 
\hline
\boldmath{$FP32_{NMNI}$}       & 79.578     & 1.675         & -9.887     & 1.506      & 99.20     \\ 
\hline
\boldmath{$FP32_{NMSI}$}       & 64.205     & 1.375         & 2.421     & 0.949      & 99.20     \\ 
\hline
\boldmath{$FP32_{NMCI}$}       & 69.832     & 1.518         & -2.367     & 1.134      & 99.20     \\ 
\hline
\boldmath{$FP32_{NMCSI}$}      & 67.095     & 1.329         & -2.088     & 1.017      & 99.20     \\
\hline
\end{tabular}
}%
\end{table}

\section{Approximate Convolutional Neural Network}
\label{sec:apprx_cnn}

\begin{figure}[tb]
    \centering
    \subfigure[]{\includegraphics[width=0.24\textwidth]{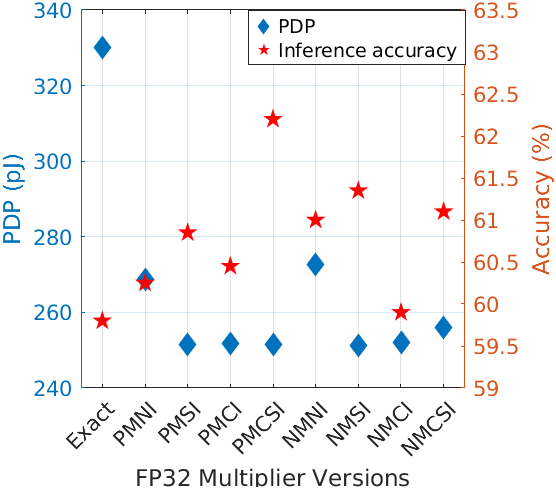}}
    \subfigure[]{\includegraphics[width=0.24\textwidth]{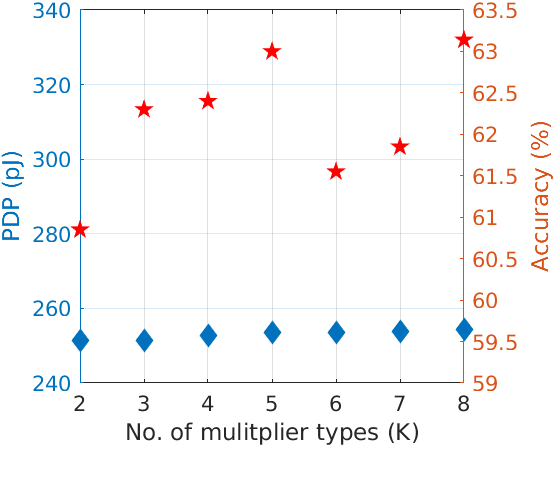}}
    \caption{PDP and CNN inference accuracy for: (a) a single multiplier across all kernels, and (b) `K' multiplier types assigned per NSGA-II-optimized sequence across convolutional layers.}
    \label{fig:pdp_acc_plots}
\end{figure}

CNN is a deep learning model that efficiently extracts spatial features from grid-like data using convolution, pooling, and nonlinear activation layers.
The framework in \cite{Bindu_metaheuristic-cnn-approx-mult} proposes layer-wise deployment of different AMs across convolutional layers using optimization algorithms, but it does not explore the flexibility of using multiple multipliers within a single layer or kernel.
In this work, a custom CNN with two convolutional layers - 10 kernels in the first and 12 in the second, each of size $3\times3$, was designed with flexible multiplier integration. Trained on CIFAR-10, the model achieved 77\% training accuracy.
For evaluation, 2000 images from the test dataset were used 
for CNN inference, where the exact FP32 multiplier achieved 59.80\% accuracy on the classification task.
Each of the 8 FP32 AMs was individually applied across all convolutional layers, with exact multipliers used elsewhere, to evaluate their impact on CNN inference. To assess hardware contribution, the power, delay, and PDP of each multiplier were linearly scaled by the total number and size of filters across all layers, while area was assumed constant, given that all multipliers are pre-implemented and reusable.
Figure \ref{fig:pdp_acc_plots}(a) shows the cumulative PDP of multipliers used in the convolutional layers alongside the resulting inference accuracy.
Most approximate multipliers outperformed the exact multiplier in inference accuracy, leveraging the error-resilient nature of CNNs. AMs introduce small, controlled noise into computations, which acts like regularization, thus reducing overfitting and improving generalization to unseen data. 

\subsection{Optimization-Based Multiplier Sequence Generation} 
\label{sec:optimization}
\begin{figure}[h]
    \centering
    \includegraphics[width=0.45\textwidth]{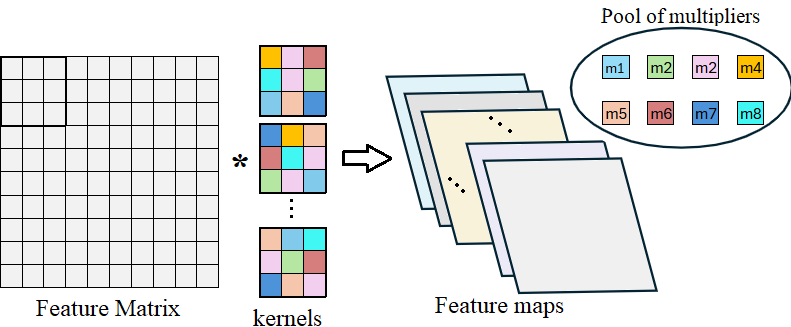}
    \caption{Representation of the convolution operation within a layer, using multiplier-interleaved kernels. }
    \label{fig:conv_representation}
\end{figure}
Figure \ref{fig:conv_representation} illustrates the convolution operation between the feature matrix and AM-interleaved kernels within a convolutional layer. From the pool of 8 AMs discussed in Section \ref{sec:FP32muls_char}, the exact multipliers in each convolution are replaced by a predefined sequence of AMs.
Each approximate multiplier is ranked from 1 to 8 based on its CNN inference accuracy when uniformly applied across convolutional layers. As shown in Figure \ref{fig:pdp_acc_plots}(a), the rankings (from highest to lowest accuracy) are: PMCSI, NMSI, NMCSI, NMNI, PMSI, PMCI, PMNI, and NMCI.
The CNN model has 22 filters (each $3\times3$), yielding 198 multiplier slots. With `K' multiplier types, the sequence space grows to $K^{198}$, making exhaustive search infeasible. To address this, the NSGA-II algorithm 
is used to optimize the multiplier sequence, minimizing area, PDP, and CNN accuracy loss.
This optimization was performed for varying numbers of multiplier types (K = 2 to 8).
\begin{figure*}[tb]
    \centering
    \subfigure[K=3]{\includegraphics[width=0.32\textwidth]{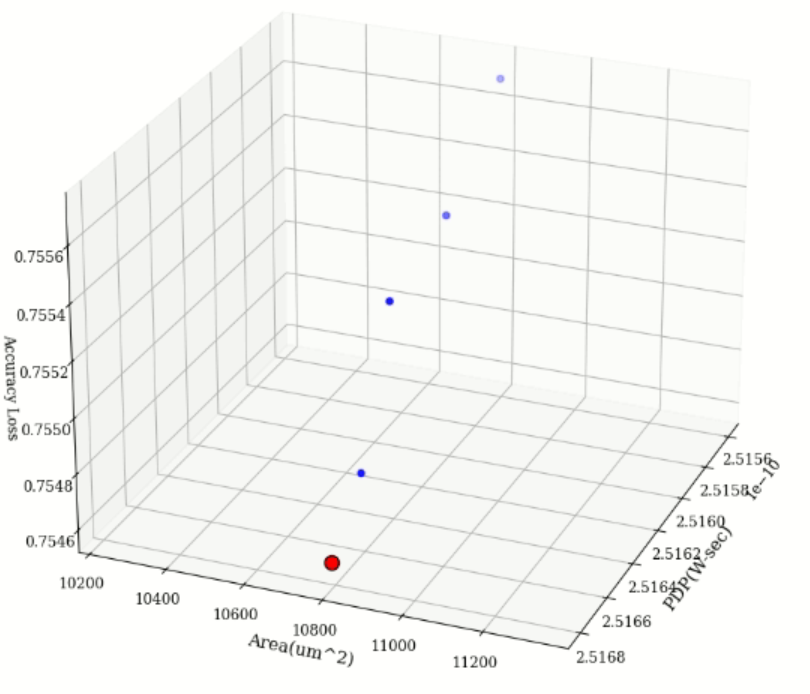}} \hfill
    \subfigure[K=6]{\includegraphics[width=0.32\textwidth]{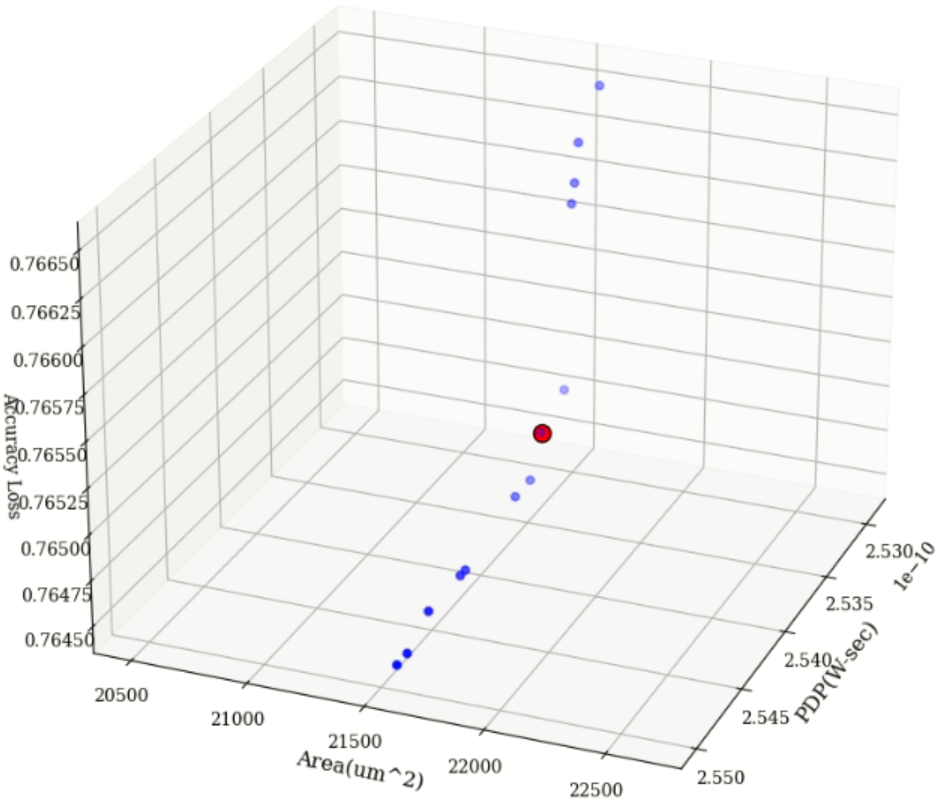}} \hfill
    \subfigure[K=8]{\includegraphics[width=0.32\textwidth]{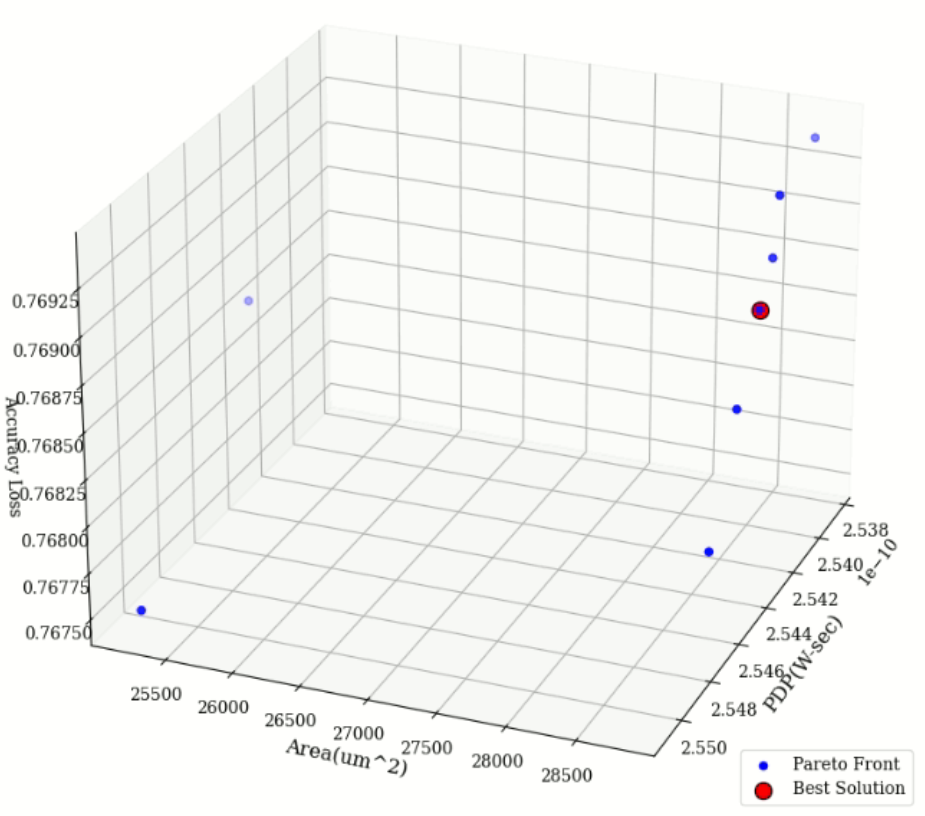}} \hfill
    \caption{Pareto-optimal solutions resulting from NSGA-II algorithm for three different 'K' values.} 
    \label{fig:NSGA2_plots}
\end{figure*}
Figure \ref{fig:NSGA2_plots} shows the 3D Pareto-optimal solutions for the considered multi-objective optimization. The solution highlighted in red from each plot was selected for CNN inference evaluation.
\begin{figure}[tb]
    \centering
    \includegraphics[width=0.36\textwidth]{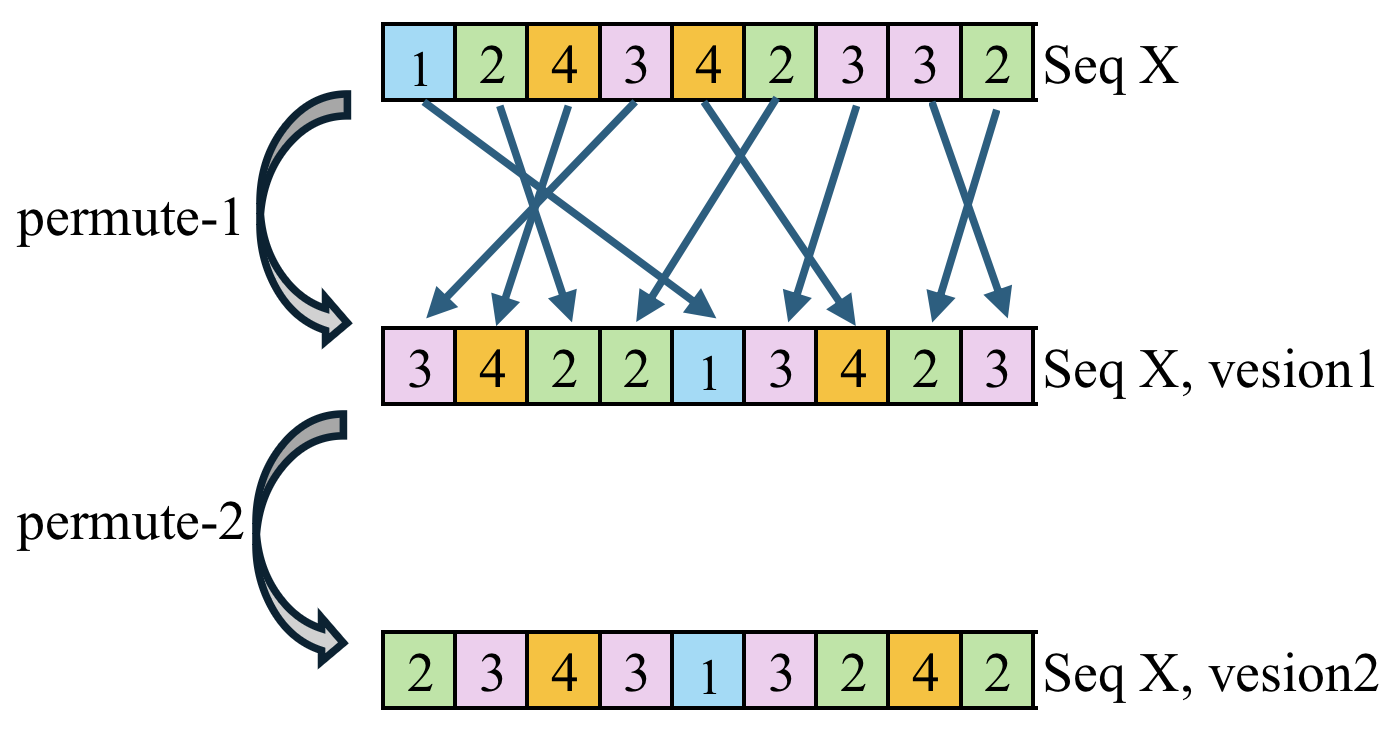}
    \caption{Illustration of randomly displaced multipliers within the optimal sequence, shown for a simplified case with 9 slots (Seq X) and K = 4 multiplier types.}
    \label{fig:permuted_versions}
\end{figure}

Since the 198 multiplier slots are unranked, the NSGA-II-generated sequence is an approximate solution. The placement order is not considered, and swapping multiplier positions does not affect overall hardware metrics. Thus, the proposed CNN optimization adopts a double-approximation strategy: first, using AMs, and second, approximating through the algorithm by ignoring positional effects within the sequence. The NSGA-II applied to extract the sequence is a faster method over NSGA-II when applied on individual slots of 198 multipliers. 
To account for this positional effects, the best NSGA-II sequence is randomly permuted across the 198 slots to generate 10 variations per `K', preserving the same set of multipliers. Figure \ref{fig:permuted_versions} illustrates this using a smaller example with 9 slots and `K'=4, 
extended to all 198-slot sequences.
The maximum CNN accuracy observed among these 10 sequence variants and the PDP contribution of the multipliers within them are reported in Figure \ref{fig:pdp_acc_plots} for all considered `K'. Notably, all the displaced sequences exhibited classification accuracy higher than the exact multiplier.
While the sequence generated for `K'=8 achieved the highest accuracy of 63.14\%, 
`K'=5, 4, and 3 yielded accuracies of 63\%, 62.4\%, and 62.3\%, respectively—each outperforming the use of a single multiplier uniformly across all convolutional layers.
The AMs selected from the NSGA-II run provide better hardware efficiency and improved model accuracy, yielding a design that is both resource-efficient and precision-enhanced.

\section{Conclusion}\label{sec:conclusion}

This work explores the optimization of CNN performance by integrating approximate FP32 multipliers within convolutional layers in a custom CNN model with two convolutional layers, allowing flexibility in multiplier selection. Eight AMs, designed by approximating the mantissa using error-diluted compressors, achieved up to 29.34\% improvement in area-power-delay product and maintained high accuracy, with 99.2\% prediction accuracy under 1\% error tolerance.
Instead of using a single multiplier type, this study applied multiple approximate FP32 multipliers in an optimized sequence across all kernel multiplications, determined using NSGA-II algorithm to balance hardware metrics like PDP and area while minimizing accuracy loss. 
Experimental results demonstrated that interleaving multipliers yielded higher classification accuracy than using a single multiplier designed uniformly across the layers, besides attaining hardware gains.
NSGA-II algorithm was run for sequence of 198 multipliers and not for specific multiplier slots in the kernel, hence the approximated sequence of multipliers were further randomly displaced and accuracy improvement was found to be consistent across.
The work establishes a double approximation scheme - i) employing inexact FP multiplier, and ii) approximating the solution to a sequence of multipliers for 22 kernels of 9 coefficients each, yet achieves improved accuracy results with hardware efficiency. The error-balancing nature of approximate multipliers enabled them to outperform the exact one in CNN inference, with controlled noise acting as regularization to reduce overfitting and improve generalization. 
The proposed framework demonstrates the usage of hardware-efficient multipliers for convolution operations to elevate CNN model performance.
\balance
\bibliographystyle{IEEEtran}
\bibliography{main}

\end{document}